\documentclass[print,showpacs,twocolumn, noshowkeys, superscriptaddress]{revtex4}%
\usepackage{amsfonts}
\usepackage{amsmath}
\usepackage{amssymb}
\usepackage{graphicx}%
\begin{document}
\title{Efficient Phase-Encoding Quantum Key Generation with Narrow-Band Single Photons}
\author{Yan Hui}
\affiliation{Laboratory of Quantum Information Technology, ICMP and SPTE, South China
Normal University, Guangzhou 510006, China}
\affiliation{Department of Physics, The Hong Kong University of Science and Technology,
Clear Water Bay, Kowloon, Hong Kong, China}
\author{Zhu Shi-liang}
\affiliation{Laboratory of Quantum Information Technology, ICMP and SPTE, South China
Normal University, Guangzhou 510006, China}
\author{Du Shengwang}
\affiliation{Department of Physics, The Hong Kong University of Science and Technology,
Clear Water Bay, Kowloon, Hong Kong, China}
\date{\today}

\begin{abstract}
We propose an efficient phase-encoding quantum secret key generation scheme
with heralded narrow-band single photons. The key information is carried by
the phase modulation directly on the single-photon temporal waveform. We show that, when the
technique is applied to the conventional single photon phase-encoding BB84 and
differential phase shift (DPS) quantum key distribution schemes, the key
generation efficiencies can be improved by a factor of 2 and 3, respectively.
For N($\geq$ 3)-period DPS systems, the key generation efficiency can be
improved by a factor of N. The technique is suitable for quantum-memory-based long-distance fiber communication system.
\end{abstract}

\maketitle

Quantum key distribution (QKD) is an unconditionally secure method to
distribute secret keys between two parties (Alice and Bob). The security of
QKD is guaranteed by the principles of quantum mechanics \cite{1,2}, such as
noncloning theorem and Heisenberg uncertainty. Since the first QKD experiment using a 32cm free-space transmission line was reported in 1992\cite{BB84}, the key distribution distance has continued to increase. With the fiber-based decoy-state BB84 protocol, a photon number splitting(PNS) secure key distribution over 200km has been achieved \cite{4}. With the differential phase shift(DPS) QKD scheme, the PNS-secure key distribution distance record is also 200km \cite{5}. The attenuated laser is worked as the source in the above schemes. In order to increase the key distribution distance, quantum memory and quantum repeater are proposed and demonstrated recently\cite{7}. And hence, single photon especially narrowband single photon is regarded as an attractive source  for long distance QKD again besides the attenuated laser\cite{6}. With single photons, phase-encoding BB84 (PE-BB84) \cite{PE-BB84-1} and the DPS-QKD \cite{DPS-1, DPS-2, DPS-3} are two typical schemes: Alice divides the single photon into two or more time slots and Bob detects the single photon using an unbalanced Mach-Zehnder (M-Z) interferometer, respectively. Because the sequenced single-photon pulses experience the same phase and polarization changes during propagation through the fiber transmission line\cite{DPS-1,DPS-2,10}, the bit error can be easily corrected at the receiver. However, due to lack of generating single photons with controllable (phase-amplitude) waveforms, in the conventional single photon PE-BB84 and
DPS-QKD schemes, a single photon is splitted into paths with different
lengthes and then recombined with passive beam splitters that introduce
unavoidable loss. As a result, the generation efficiency decreases as the
number of time slot increases. In order to increase the generation efficiency, many methods have been proposed, such as using optical switches or polarization beam splitter\cite{11}.

\begin{figure}
[ptb]
\begin{center}
\includegraphics[
height=6.2296cm,
width=7.9452cm
]%
{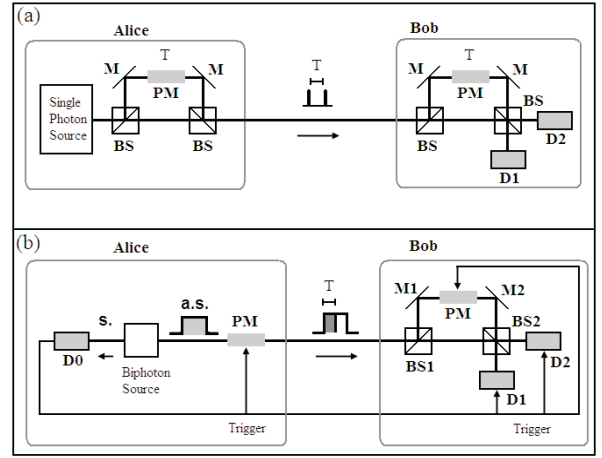}%
\caption{(a) the conventional phase-encoding BB84 scheme; (b) our proposed
phase-encoding BB84 scheme with heralded narrow bandwidth square wave single photons;
s.: Stokes; a.s.: anti-Stokes; PM: phase modulator; BS, BS1, BS2: 50\% beam splitter; M, M1, M2: mirror; D, D1, D2: single-photon detector.}%
\label{fig2}%
\end{center}
\end{figure}

In this paper, we propose another phase-encoding generation method to improve the
key creation efficiency in the PE-BB84 and DPS-QKD schemes without using M-Z interferometer on Alice's site. The motivation comes from the recent
narrow-band nonclassical paired photon generation \cite{Balic2005,
Vuletic2006, Lukin2003, Kimble2003}. Using spontaneous four-wave mixing and
electromagnetically induced transparency in cold atoms, subnatural linewidth
biphoton with a coherence time up to about 1 $\mu s$ has been demonstrated
\cite{subnatural}. Du \textit{et al.} proposed and demonstrated shaping
biphoton temporal waveforms by periodically modulating the two classical
driven fields \cite{Du2010}. With such a long coherence time and under
detecting one of the paired photons, heralded single photons with arbitrary
phase-amplitude waveform can be generated with external light modulators
\cite{EOModulator, 14}. It is then possible to eliminate the need of beam
splitters in the conventional phase-encoding schemes by using directly phase
modulated heralded narrow-band single photons.

%



%

\begin{figure}
[ptb]
\begin{center}
\includegraphics[
height=5.452cm,
width=5.5289cm
]%
{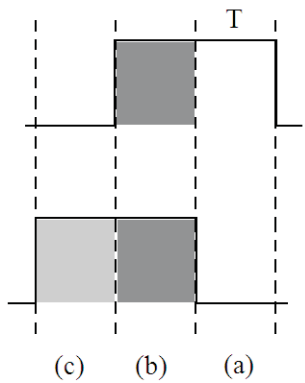}%
\caption{Time squences of the phase modulated square wave single photon of the
proposed phase-encoding BB84 scheme, the ratio of the click probabilty in the
three time squences is $(a):(b):(c)=1:2:1$. $T$ is the phase modulated
period(Here we suppose $T=100$ $ns$).}%
\label{fig3}%
\end{center}
\end{figure}

We first consider the PE-BB84 scheme and improve its key generation
efficiency. Figure \ref{fig2}(a) shows the conventional setup. On Alice's
site, a single photon is divided into one short path and one long path with
phase modulation (PM) and a time delay of $T$ after the first beam splitter
(50\%), and then is recombined at the second beam splitter (50\%). This
effectively splits the single photon as a superposition of two time slots
separated by time T. The phase difference between these two time slots is
modulated by one of the two nonorthogonal basis $\{0,\pi\}$ and $\{\pi/2,3\pi/2\}$
randomly. Bob measures the phase difference with two detectors either in the
$\{0,\pi\}$ or the $\{\pi/2,3\pi/2\}$ basis, using a phase modulator in the
long path of an unbalanced M-Z interferometer whose path difference equals to
that on Alice's site. It is clear that the single photon at Alice's site has
50\% probability leaking out of the system and thus the maximum key sending
efficiency is $1/2$. On Bob's site, there is no photon loss through the beam
splitters. To maximize the use efficiency of the single-photon source, we
better avoid the beam splitters on Alice's site. Our proposed scheme is shown
in Fig.~\ref{fig2}(b) where Alice's site is modified. In our scheme, with the
technique described in \cite{EOModulator, 14} and feedback waveform control, Alice makes use of narrow-band biphotons to generate heralded single anti-Stokes photon with a rectangular shape with a temporal
length of 2T (for example, T=100 ns). This rectangular-shaped single photon
then passes through a PM trigged by detection of the Stokes
photon and the phase difference is encoded to the two time slots. The
detection at Bob's site is similar to that in the conventional scheme
[Fig.~\ref{fig2}(a)] except the trigger timing of detecting the Stokes photon is sent from Alice through a classical channel. In this way, there is no photon loss on Alice's site and
the key generation efficiency is increased by a factor of 2. Bob could detect
one photon at the three time instances with the ratio $1:2:1$ as illustrated
in Fig.\ref{fig3}: (a) the first period of the photon passes the short path of
the M-Z interferometer; (b)the first period of the photon passes the long path
and the second period of the photon passes the short path of the M-Z
interferometer; (c)the second period of the photon passes the long path of the
M-Z interferometer. In the time instance (b), the phase difference between the
two consecutive periods will determine the outputs of the M-Z interferometer
and then the click of which detector. When Bob detects a photon at the time
instances (a) and (c), he will discard data. When Bob detects a photon at the
time instance (b), a secret key bit can be created by comparing his basis with
Alice's, similar to the protocol in polarization-based BB84 \cite{BB84}.
Because Bob has 1/2 probability in measuring the phase difference and another
1/2 probability in matching the basis, the receiving-key efficiency is 1/4.
Therefore accounting the sending efficiency on Alice's side, the total key
creation efficiency is 1/8 for the conventional PE-BB84 scheme, and 1/4 for
the improved scheme. The security of the PE-BB84 scheme has been analyzed a
lot in the past and proven to be unconditionally secure \cite{2,16}.

%

\begin{figure}
[ptb]
\begin{center}
\includegraphics[
height=6.267cm,
width=7.9496cm
]%
{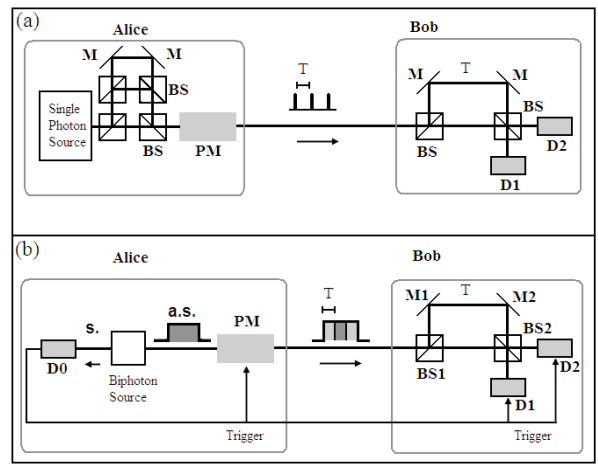}%
\caption{(a) the currently implemented DPS-QKD scheme; (b) our proposed
DPS-QKD scheme with heralded narrow bandwidth square wave single photons; s.: Stokes; a.s.: anti-Stokes; PM: phase
modulator; BS, BS1, BS2: 50\% beam splitter; M, M1, M2: mirror; D, D1, D2: single-photon detector.}%
\label{fig4}%
\end{center}
\end{figure}
\begin{figure}
[ptbptb]
\begin{center}
\includegraphics[
height=5.485cm,
width=8.0133cm
]%
{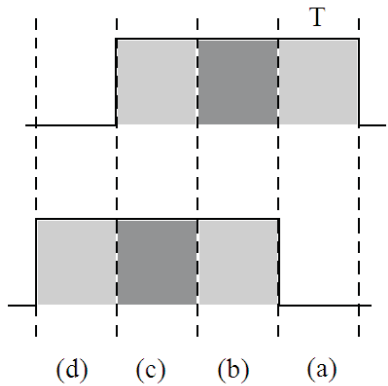}%
\caption{Time squences of the phase modulated square wave photon of our
proposed DPS-QKD scheme, the ratio of the click probabilty in the four time
squences is $(a):(b):(c):(d)=1:2:2:1$. $T$ is the phase modulated period(Here
we suppose $T=100$ $ns$).}%
\label{fig5}%
\end{center}
\end{figure}

Now we turn to the DPS-QKD scheme that has been demonstrated to be one of the most
applicable schemes\cite{PE-BB84-1}. Here we show that with a phase
modulated long single photon from a biphoton source, the key creation efficiency of the DPS-QKD
scheme could also be improved significantly. Figure \ref{fig4}(a) shows the
setup of the conventional DPS-QKD scheme \cite{DPS-1}. On Alice's site, the
photon from a single photon source is divided into three paths with time
separation $T$ and then recombined by beam splitters. The keys are encoded by
preparing the relative phase shift between two consecutive pulses in $0$ or
$\pi$ randomly. Bob measures the phase difference using an unbalanced M-Z
interferometer setup with a path difference that compensates the time delay T.
Similar to that in PE-BB84, the sending efficiency of a DPS photon is only 1/3
due to the loss of beam splitters. Such a loss can be eliminated in our
improved scheme without using beam splitters on Alice's site, as shown in
Fig.\ref{fig4}(b). Similar to the improved PE-BB84 system, we divide the long
rectangular-shape photon with a temporal length 3T into 3 time sequences with
equal period T. As one does not know the exactly arriving time of the single
photon within the three time slots, the heralded single photon can be
described as a superposition of $|1_{a}0_{b}0_{c}\rangle$, $|0_{a}1_{b}0_{c}\rangle$, and $|0_{a}0_{b}1_{c}\rangle$ (where "$1_a$" represents the photon at time slot $a$, otherwise it is
"$0_a$"). Because the phase of each time slot is randomly modulated by $0$ or
$\pi$, the photon sent from Alice to Bob is in one of the four
states: $1/\sqrt{3}(|1_{a}0_{b}0_{c}\rangle \pm |0_{a}1_{b}0_{c}\rangle \pm |0_{a}0_{b}1_{c}\rangle)$ in the present scheme. These four states, which are nonorthogonal with each other and thus cannot be identified by a single measurement, have the same mathematica forms as those in the conventional DPS-QKD scheme \cite{DPS-1, DPS-3}. Therefore, the unconditional security of the proposed scheme can be proved following the procedure in Ref.~\cite{DPS-3}. The detection setup on Bob's site is similar to that in the conventional scheme with the trigger timing sent from Alice through a classical channel by detecting the Stokes photons. It is clear that the single-photon sending efficiency
becomes unity in this case and the encoding machine is lossless. In the
DPS-QKD configuration, Bob detects a photon at four possible time instances
with the ratio $1:2:2:1$ as illustrated in Fig. \ref{fig5}: (a) the photon in
the first period passes the short path of the M-Z interferometer; (b) the
photon in the first period passes the long path and the photon in the second
period passes the short path; (c) the photon in the second period passes the
long path and the photon in the third period passes the short path; (d)the
photon in the third period passes the long path. In the time instances (b) and
(c), the phase difference between the proper consecutive periods will
determine the outputs of the M-Z interferometer and then the click of which
detector. Bob discards the photons detected at the time instances (a) and (d),
and communicate with Alice the time instance when he get a photon click only
at (b) or (c) \cite{DPS-1}. With her own modulation pattern, Alice knows which
detector clicked on Bob's site and key bits are created and shared by the two
parties. The details of the protocol can be seen in Ref.\cite{DPS-1}. Here we focus on the
key creation efficiency. On Bob's site, photons counted at the time instances
(b) and (c) fully contribute to the key. The probability for these events is
$2/3$. Thus, taking into account the sending efficiency on Alice's site, the
entire key creation efficiency is $2/9$ for the conventional
beam-splitter-based DPS-QKD scheme, and 2/3 for our improved scheme. Another
feature that should be mentioned is the information capacity after error
correction. As described above, the efficiency to obtain sifted keys in our
scheme is $3$ times that in the conventional DPS-QKD scheme, while the error
rate introduced by the simple intercept/resend attack is $1/4$, which is the
same as the other scheme. Thus, the new scheme has the larger final
information capacity when other parameters hold the same.%

\begin{figure}
[ptb]
\begin{center}
\includegraphics[
height=6.0868cm,
width=7.9496cm
]%
{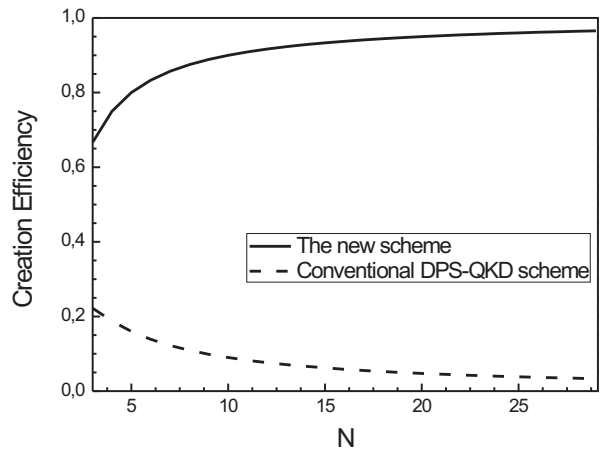}%
\caption{The key creation efficiency with different N for our proposed scheme
and the conventional DPS-QKD scheme(with passive BS).}%
\label{fig6}%
\end{center}
\end{figure}

As proposed by Inoue \textit{et al.} \cite{DPS-1,19}, the above DPS-QKD scheme
with N=3 time slots can be extended to N($>$3) cases where the key receiving
efficiency scales as (N-1)/N and approaches 1 at large N limit. However, in
the conventional setup with passive beam splitters on Alice's site, the
single-photon sending efficiency decreases at a larger N because it scales as
1/N. As a result, the total key creation efficiency becomes (N-1)/N$^{2}$ and
decreases to zero at the limit of large N. If we use heralded narrow-band
single photons with proper phase-amplitude modulations, the sending efficiency
on Alice's site is always 1 and does not depend on N. Thus, in our proposed
technique, the total key creation efficiency is proportional to (N-1)/N and
indeed reaches unity at large N limit. Figure \ref{fig6} shows the difference
of the total creation efficiencies as functions of N between the proposed and
the conventional DPS-QKD schemes as a comparison.

In addition, besides the key creation effeciency, the secure key rate(SKR) is
another important parameter to characterize a practical QKD system. Comparing with the conventional single photon schemes, there are several more parameters which will limit the SKR in our scheme: Firstly, the generation rate and the temporal length of the heralded single photon; with 300ns
temporal length photons, the SKR will be limited to about 3MHz; using faster phase
modulator($>$ 30 GHz) allows us to reduce the temporal length from 300ns to
several ns or even shorter\cite{20}; the generation rate can be increased if we short the temporal length of the single photon\cite{EOModulator}, or using spontaneous parametric down-conversion heralded single photon source. Secondly, the time jitter of the detector, with the time jitter of 100ps, the detector will bring a error rate of 1\% if T=10ns. Thirdly, the shape of the square wave single photon after a long distance transmission, especially the rising and falling edges, this shortage can be conquered if we let the temporal length of the single-photon a little longer than NT(the effective signal length); the propagation losses in the optical fiber cables is bigger for the 780nm narrow-band single photons as we proposed; fortunately, the generation of telecom wavelengths narrow-band photons has already been demonstrated in experiment\cite{21}.

In summary, we have proposed a high efficient phase-encoding quantum key
generation scheme by using heralded narrow-band single photons with phase
modulation. While implemented to the single photon PE-BB84 protocol, the entire key creation
efficiency can be increased by a factor of 2. For the single photon DPS-QKD scheme with N=3,
the key creation efficiency can be increased by a factor of 3. We further show
that in the conventional single photon scheme the entire key creation efficiency decreases
as we increase N and reaches zero at large N limit due to the beam splitter
loss on Alice's site. In our proposed technique, the creation efficiency
scales as (N-1)/N and reaches unity at large N. The overall maximum efficiency
may be only limited by shaping loss of the initial temporal waveform of
heralded single photons emitted from their source, the quantum detection
efficiencies, and propagation loss. The nearly rectangular-shaped subnatural
linewidth biphotons \cite{subnatural} are ideal for this application with a
reshaping loss less than 20\%. In addition, the narrow band photons can be directly integrated with the quantum memory and quantum repeater, so the described technique will be suitable for quantum-memory-based
long-distance fiber transmission systems.\bigskip

\begin{acknowledgments}
The authors acknowledge helpful discussions with He GP and Wang XB. The work was supported by the Hong Kong Research Grants
Council (Project No. HKUST600809). HY and SLZ were aslo supported
by the NSF of China under Grant No. 10974059 and the State Key
Program for Basic Research of China (Grants No. 2011CB922104 and
No. 2007CB925204).
\end{acknowledgments}


\begin{thebibliography}{99}                                                                                               %


\bibitem {1}Gisin N, et al.,2002 Rev. Mod. Phys.
\textbf{74}, 145

\bibitem {2}Scarani V, et al.,2009 Rev. Mod. Phys. \textbf{81}, 1301

\bibitem {BB84}Bennett C H  and Brassard G,1984 in Proceedings of the IEEE
International Conference on Computers, Systems, and Signal Processing,
Bangalore, India (IEEE, New York), p. 175

\bibitem {4}Liu Yang et al.,2010 Opt. Express 18, 8587; Rosenberg D, et
al.,2009 New J. Phys. 11, 045009; Wang X B,2005 Phys. Rev. Lett. 94,
230503; Lo H K, et al.,2005 Phys. Rev. Lett. 94, 230504

\bibitem {5}Takesue H, et al.,2007 Nature Photonics 1, 343

\bibitem {7}Tanji H, et al.,2009 Phys. Rev. Lett. \textbf{103}, 043601;Duan L, M,  et al.,2001 Nature \textbf{414}, 413; Jiang L, et al.,2007 Phys. Rev. A \textbf{76}, 012301;Chen Z B, et al.,2007 Phys. Rev. A \textbf{76}, 022329

\bibitem {6}Waks E, et al., 2002 Nature 420, 762;Beveratos A, et al., 2002 Phys.
Rev. Lett. 89, 187901

\bibitem {PE-BB84-1}Bennett C H,1992 Phys. Rev. Lett. \textbf{68}, 3121;
Marand C  and Townsend P D, 1995 Opt. Lett. \textbf{20}, 1695

\bibitem {DPS-1}Inoue K,et al., 2002 Phys. Rev. Lett.
\textbf{89}, 037902

\bibitem {DPS-2}Inoue K,et al., 2003 Phys. Rev. A \textbf{68},
022317

\bibitem {DPS-3}Wen K,et al., 2009 Phys. Rev. Lett.
\textbf{103}, 170503

\bibitem {10}Chen X, et al.,2004 Appl. Phys. Lett.\textbf{85}, 1648

\bibitem {11}Adachi et al.,2009 New J. Phys. \textbf{11}, 113033;Lo H K, Chau H F , and
Ardehali M,2005 J. Cryptology \textbf{18}, 133; Gobby C, Yuan Z L, and Shields A J,2004 Appl. Phys. Lett. \textbf{84}, 3762

\bibitem {Balic2005}Balic V,et al.,2005 Phys. Rev. Lett. \textbf{94}, 183601; Kolchin P, et al.,2006 Phys. Rev. Lett. \textbf{97}, 113602

\bibitem {Vuletic2006}Thompson J K,et al.,2006,
Science \textbf{313}, 74

\bibitem {Lukin2003}van der Wal C H,et al.,2003 Science
\textbf{301},196

\bibitem {Kimble2003}Kuzmich A,et al.,2003 Nature(London) \textbf{423}, 731

\bibitem {subnatural}Du S W,et al.,2008 Phys. Rev. Lett. \textbf{100}, 183603

\bibitem {Du2010}Du S,et al.,2009 Phys. Rev. A \textbf{79},
043811; Chen J F,et al.,2010 Phys. Rev. Lett. \textbf{104}, 183604

\bibitem {EOModulator}Klchin P, et al.,2008 Phys. Rev. Lett. \textbf{101}, 103601

\bibitem {14}Specht H P, et al.,2009 Nature Photonics \textbf{3}, 469

\bibitem {16}Mayers D, 1996 in Advances in Cryptology: Proceedings of Crypto96,
Lecture Notes in Computer Science Vol. 1109 (Springer-Verlag, Berlin),
p. 343; Shor P W  and Preskill J,2000 Phys. Rev. Lett. \textbf{85}, 441

\bibitem {19}Zhou C,et al.,2003 Appl. Phys. Lett.
\textbf{83}, 1692

\bibitem {20}Chen J F,et al.,2010 Phys. Rev. Lett. \textbf{104}, 223602

\bibitem{21}Chaneliere T,et al., 2006 Phys. Rev. Lett. \textbf{96}, 093604

\end{thebibliography}
\end{document}